\documentclass[reprint,aps,prb,longbibliography,showpacs,amsmath,amssymb,superscriptaddress]{revtex4-2}
\usepackage[english]{babel}
\usepackage{amsmath,amssymb,amsfonts} 
\usepackage{graphicx}
\usepackage[utf8]{inputenc}
\usepackage{bm}
\usepackage{xcolor}
\usepackage[normalem]{ulem}
\usepackage{SIunits}
\usepackage[colorlinks=true,linkcolor=blue,citecolor=blue,urlcolor=blue]{hyperref}
\usepackage{url}
\usepackage{textcomp}
\usepackage{lineno,hyperref}
\usepackage{float}
\usepackage{chemfig}
\usepackage{rotating}
\usepackage{subcaption}
\usepackage{color} 
\usepackage{color} 
\definecolor{red}{rgb}{1.,0.,0.}
\usepackage{soul}
\usepackage{amsmath}

\usepackage{graphicx,wrapfig,lipsum}
\usepackage{caption}
\DeclareUnicodeCharacter{00A0}{~}
\usepackage{xcolor}  

\makeatletter
\newcommand*{\rom}[1]{\expandafter\@slowromancap\romannumeral #1@}
\makeatother

\usepackage[inline]{asymptote}

\begin{asydef}
// Global Asymptote definitions
real linkLen=1, linkWidth=2pt;
real rl=2+linkLen;              // distance between beads
guide link=(1,0)--(1+linkLen,0);   // a link
pen beadColor=orange;
pen linkColor=beadColor;
void bead(transform t){
  draw(t*link,linkColor+linkWidth);
  radialshade(t*unitcircle,
    beadColor,shift(t)*(-0.4,0.3),0.01
   ,black,shift(t)*(-0.4,0.3),1.5);
}
pair operator>(pair pos=(0,0), real phi){
  transform t=shift(pos)*rotate(phi);
  bead(t); // draw a bead with a link
  pos+=rl*(Cos(phi),Sin(phi)); // Sin, Cos - in degrees, sin, cos - in radians
  return pos;
};
pair pos;
\end{asydef}
\usepackage{tikz}
\usepackage[labelfont=bf,
justification=raggedright]{caption}
\newcommand\HARVARD{John A. Paulson School of Engineering and Applied Sciences,
Harvard University, Cambridge, MA 02138, USA}

\newcommand\HARVARDPHYS{Department of Physics, Harvard University, Cambridge, Massachusetts 02138, USA}
\newcommand\OXFORD{Department of Materials, University of Oxford, Parks Road, Oxford OX1 3PH, United Kingdom}

\newcommand\AALTO{Department of Applied Physics, Aalto University, 02150 Espoo, Finland}

\newcommand\IMPERIAL{Departments of Physics and Materials and the Thomas Young Center for Theory and Simulation of Materials, Imperial College London, London SW7 2AZ, United Kingdom}

\graphicspath{{./images_paper/}}

\newcommand{\editor}[2]{%
\expandafter\newcommand\csname #1cancel\endcsname[1]{%
    \textcolor{#2}{\sout{##1}}}%
  \expandafter\newcommand\csname #1note\endcsname[1]{%
    \textcolor{#2}{(\textbf{#1:} ##1)}}%
  \expandafter\newcommand\csname #1\endcsname[1]{%
    \textcolor{#2}{##1}}%
  \expandafter\newcommand\csname #1change\endcsname[2]{%
    \textcolor{#2}{\sout{##1} ##2}}%
  \newenvironment{#1text}{\color{#2}}{\color{black}}
}

\editor{nr}{red}

\begin{document}

\title{Exploring Charge Density Waves in two-dimensional NbSe$_2$ with Machine Learning}

\author{Norma Rivano}\thanks{Corresponding author: nrivano@g.harvard.edu} 
\affiliation{\HARVARD}

\author{Francesco Libbi}
\affiliation{\HARVARD}

\author{Chuin Wei Tan}
\affiliation{\HARVARD}

\author{Christopher Cheung}
\affiliation{\IMPERIAL}

\author{Jose L. Lado}
\affiliation{\AALTO}

\author{Arash Mostofi}
\affiliation{\IMPERIAL}

\author{Philip Kim}
\affiliation{\HARVARDPHYS}

\author{Johannes Lischner}
\affiliation{\IMPERIAL}

\author{Adolfo O. Fumega}\thanks{Corresponding author: adolfo.oterofumega@aalto.fi} 
\affiliation{\IMPERIAL}
\affiliation{\AALTO}

\author{Boris Kozinsky}\thanks{Corresponding author: bkoz@seas.harvard.edu}
\affiliation{\HARVARD}

\author{Zachary A. H. Goodwin}\thanks{Corresponding author: zac.goodwin@materials.ox.ac.uk} 
\affiliation{\HARVARD}
\affiliation{\IMPERIAL}
\affiliation{\OXFORD}

\begin{abstract}
Niobium diselenide (NbSe$_2$) has garnered significant attention due to the coexistence of superconductivity and charge density waves (CDWs) down to the monolayer limit.
However, realistic modeling of CDWs—capturing effects such as layer number, twist angle, and strain—remains challenging due to the high computational cost of first-principles methods.
Here, we develop a physically informed workflow for training machine-learning interatomic potentials (MLIPs) based on the E(3)-equivariant Allegro architecture, tailored to capture the subtle structural and dynamical signatures of CDWs in mono- and bilayer NbSe$_2$.
We find that while CDW lattice distortions are relatively easy to learn, modeling vibrational properties remains more challenging.
It requires targeted dataset design and careful hyperparameter tuning, pushing the boundaries and testing the extensibility of current MLIP frameworks. 
Our MLIPs enable reliable simulations of commensurate and incommensurate CDW phases, including their sensitivity to dimensionality and stacking, as well as CDW dynamics, phonons, and transition temperatures estimated via the stochastic self-consistent harmonic approximation.
This work opens new possibilities for studying and tuning CDWs in NbSe$_2$ and other two-dimensional systems, with implications for electron-phonon coupling, superconductivity, and advanced materials design.
\end{abstract}

\maketitle

\section{Introduction}

Niobium diselenide ($\mathrm{NbSe_2}$) has long been studied for its unique interplay between charge density wave (CDW) order and superconductivity, which persists even in the monolayer limit \cite{ugeda2016characterization,xi2015strongly}.
As a van der Waals crystal,
the bulk can be exfoliated into mono- and multilayers, providing a versatile platform for probing dimensionality-dependent phenomena.~\cite{lian2017first,zhang2022tailored,lin2020patterns,nakata2018anisotropic,xi2016ising, baidya2021transition, xi2015strongly, calandra2009effect, leroux2012anharmonic, bianco2020weak,Silva2016}
Unlike quasi-one-dimensional materials, where CDWs are often driven by Fermi surface nesting~\cite{peierls1930theorie,peierls1996quantum,luttinger1963exactly,pouget2016peierls}, two-dimensional (2D) $\mathrm{NbSe_2}$ lacks such features. 
Instead, electron–phonon coupling (EPC) is the driving mechanism ~\cite{nakata2018anisotropic, xi2016ising, weber2011extended, zheng2019electron}, making a clear understanding of CDWs essential both for uncovering the microscopic mechanisms of superconductivity~\cite{hamill2021two, wickramaratne2020ising, khestanova2018unusual,cho2018using,xi2016ising}  and for engineering tunable quantum phases in devices~\cite{chen2020strain,xi2016gate,li2022high, ding2025gate,dvir2021planar}.

The subtlety of electron-phonon coupling and the small energy scales involved make CDWs a continuing focus of research~\cite{zhang2022tailored, lin2020patterns, khestanova2018unusual, baidya2021transition, xi2015strongly, weber2011extended, calandra2009effect}.
While reduced dimensionality can strengthen CDWs~\cite{calandra2009effect}, it may also introduce fluctuations that disrupt long-range order~\cite{gruner2018density}.
In monolayer NbSe$_2$, several density-functional theory (DFT) studies \cite{lian2018unveiling,cossu2018unveiling,guster2019coexistence} have identified multiple  $3\times$3 CDW phases. Among these, the \lq hollow\rq ~phase, where Nb atoms displace toward the center of the hexagonal unit cell (see Fig.~\ref{fig2}a), best matches experimental measurements \cite{lian2018unveiling}.
For this phase, transition temperatures up to 145 K \cite{ugeda2016characterization,xi2015strongly} have been observed, significantly higher than the bulk value of 33 K \cite{ugeda2016characterization}, though
other studies report weaker dimensionality dependence (73 K \cite{bianco2020weak} and 25--45 K \cite{arguello2014visualizing}).
For bilayers and multilayers, CDW phases are less well understood: experiments generally favor the hollow configuration, although theory sometimes predicts alternative structures \cite{cossu2024stacking}. 
Beyond dimensionality, CDWs are strongly affected by external perturbations, including strain \cite{lian2017first}, doping \cite{lian2017first,xi2016gate}, stacking order \cite{cossu2024stacking}, and substrates \cite{dreher2021proximity}. 
This sensitivity naturally extends to twisted bilayers where moiré superlattices, created by introducing twist angles between 2D layers, provide an additional route to tune materials properties \cite{Carr2017,Cao18CI,Cao18US,kerelsky2019maximized,Hao21,park2022robust,zhang2018moire,rosenberger2020twist,mak2022semiconductor,Kennes21}.
Yet, despite their importance,  
CDWs in multilayer and moiré systems remain poorly understood and challenging to simulate with first-principles methods \cite{Kennes21,goodwin2022moire, Cheung2024}.

To address the complexity of simulating CDWs, we turn to machine learning interatomic potentials (MLIPs), a powerful alternative for simulating large systems with high fidelity at reduced computational costs~\cite{Behler21CR,Deringer19AM,Deringer21CR,Falletta24,Goodwin24}. 
While MLIPs have been developed for various 2D materials and transition metal dichalcogenides 
\cite{Rowe2018,Thiemann2020,Siddiqui2024}, their application to CDWs in 2D systems is rare.
This represents a particularly challenging case, as commensurate CDWs typically involve periodic lattice distortions ($\sim$0.1~\AA) accompanied by energy changes of the order of 1 meV/atom, testing the accuracy limits of current MLIPs.
To date, the only concurrent effort for NbSe$_2$ is the study by Benić et al. \cite{benić2025machine}, who developed an electronic free-energy MLIP for monolayers to capture electronic temperature effects and investigate CDW melting under thermal and nonthermal conditions. 

Here, we develop Allegro-based MLIPs~\cite{musaelian2023learning} that reconstruct CDW phases in monolayers and bilayers, with extensibility to incommensurate and moir\'e superstructures. We present a physically informed workflow for training MLIPs, including guidelines for dataset design, hyperparameters, and validation.
While the models are specific to NbSe$_2$, the workflow and training principles are broadly applicable to other 2D CDW systems.
We validate our approach by comparing structural and phonon properties from molecular dynamics (MD) simulations with DFT results, with special attention to commensurate and incommensurate supercells essential for modeling moir\'e systems. 
In addition, we demonstrate the sensitivity of CDWs to layer number, stacking, and incommensurability, with important implications for electron–phonon coupling and superconductivity.

\section{Results and discussion}\label{results}

\subsection{Monolayer - No CDW Phase}

\begin{figure*}
    \centering    
    \includegraphics[width=0.99\linewidth]{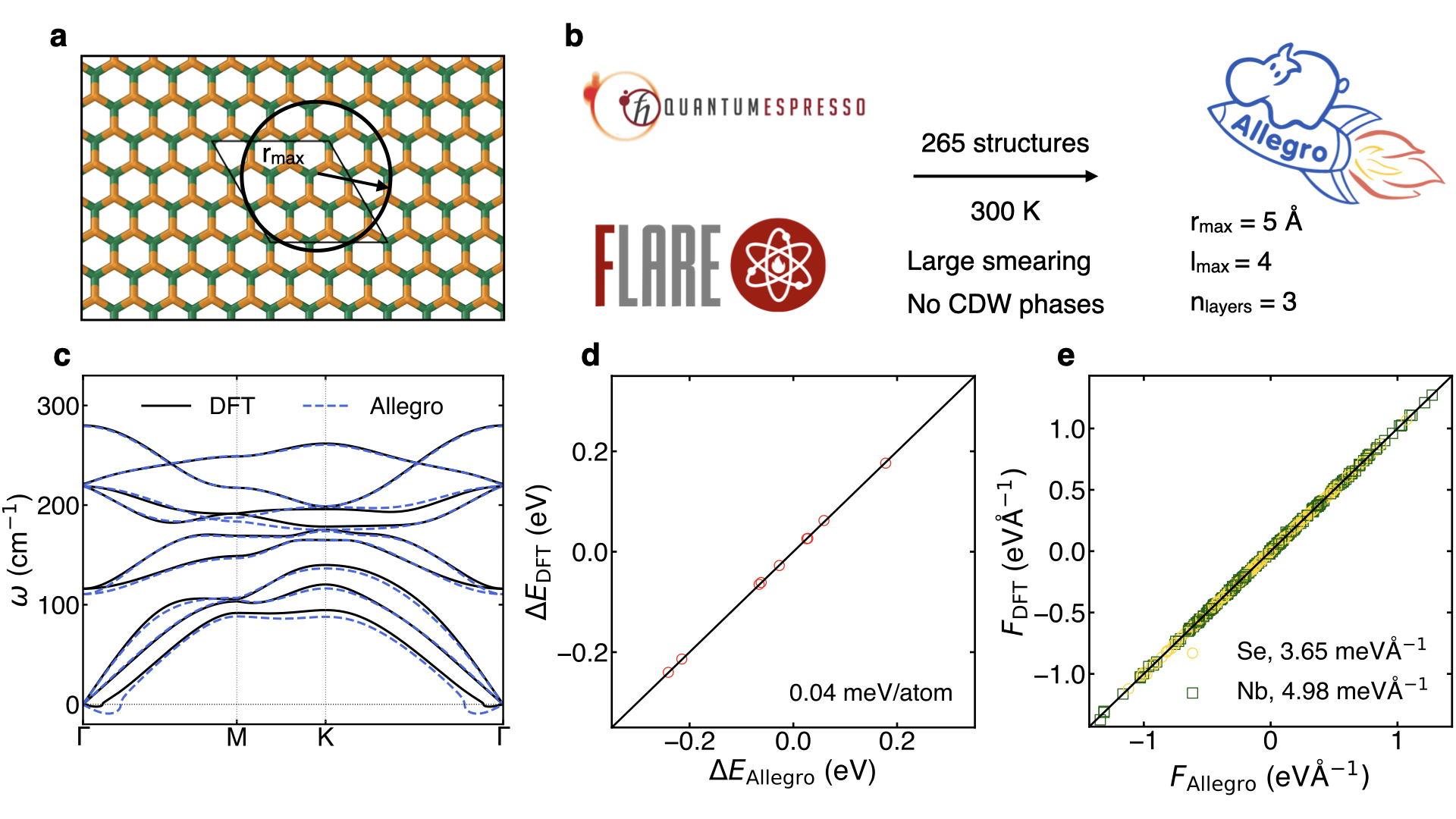}
    \caption{\textbf{Development of a machine-learned interatomic potential (MLIP) for NbSe$_2$ in the normal state (large smearing, no CDW).} 
    \textbf{a} - Structure of monolayer $\mathrm{NbSe_2}$ in the normal phase, visualized using Ovito \cite{stukowski2009visualization}, showing a $3\times3$ supercell. Orange sticks represent Se atoms, and green sticks represent Nb atoms. The range of interactions ($r_\mathrm{max} = 5$~\AA) is indicated schematically. 
    \textbf{b} - Computational workflow employed to develop the MLIP. 
    \textbf{c} - Phonon dispersion relations from DFT (solid lines) and Allegro MLIP (dashed lines), demonstrating close agreement. 
    \textbf{d} - Energy mean absolute error (MAE) per simulation frame from molecular dynamics (MD) simulations at 200~K in the NVT ensemble,  comparing Allegro MLIP predictions with DFT. 
    \textbf{e} - Corresponding force MAEs; values are reported per element as shown in the legend. Further details are given in Methods and SI.} 
    \label{fig1}
\end{figure*}

Owen \textit{et al.}~\cite{Owen2024} showed that potential energy surfaces of certain transition metals, including Nb, are very sensitive functions of atomic positions due to the sharply varying density of d-states near the Fermi energy. 
This sensitivity requires high angular resolution to capture complex many-body interactions, making these materials challenging even for state-of-the-art neural equivariant MLIPs \cite{Batzner22}.
As 2D NbSe$_2$ is  metallic  with d-orbital-dominated states near the Fermi level~\cite{lian2018unveiling}, constructing an accurate MLIP might not be trivial.
As such, we begin by training a model for the monolayer in the normal phase, using large electronic smearing (effective electronic temperature). This simplifies the energy landscape, suppresses the CDW, and stabilizes the normal state, aiding the development of a robust potential energy surface model.

In the normal state, monolayer NbSe$_2$ consists of a triangular lattice of Nb atoms bonded to six Se atoms. Se atoms form triangular lattices above and below the Nb plane, as shown in Fig.~\ref{fig1}a, where the $3\times$3 supercell used to develop our MLIP is displayed. 
Starting from this structure, we performed Bayesian active learning (BAL) using the Fast Learning of Atomistic Rare Events (FLARE) architecture \cite{vandermause2020fly} and Quantum Espresso (QE) \cite{giannozzi1991ab,giannozzi2017advanced,giannozzi2009quantum} to provide reference electronic structure calculations. We ran several BAL trajectories at 300~K, collecting 265 frames to train the model.
Further details can be found in the Supplementary information (SI) \cite{supplementary}, with a summary of the workflow in Fig.~\ref{fig1}b. 

As MLIP performance heavily depends on the hyperparameters, we conducted a hyperparameter scan (see SI). 
We found that a satisfactory value for $r_\mathrm{max}$,
the interaction length scale in the model, is 5~\AA; a good value for the models spherical harmonics resolution is $l_\mathrm{max}$ = 4; and for the number of layers ($n_\mathrm{layers}$) of the neural network, describing the complexity of the interactions included, a value of 3 is sufficient. These hyperparameters are relatively standard choices, except for $l_\mathrm{max} = 4$,  which is unusually large as required by the large
density of d-states near the Fermi energy ~\cite{Owen2024}. 

After selecting the hyperparameters, we ran MD with this
MLIP at 200~K, collecting structures, potential energies, and forces every 0.5~ps. These structures were used to compute the corresponding DFT ground-state with the resulting parity plots shown in Fig.~\ref{fig1}d-e. We find excellent agreement between our MLIP and DFT, with a force mean absolute error (MAE) of 3.65~meV$\mathrm{\AA^{-1}}$ for Se and 4.98~meV$\mathrm{\AA^{-1}}$ for Nb, corresponding to an average force error of 1.4\%, and an energy MAE of 0.04~meV/atom.

We then tested our model by comparing the phonon dispersions from the trained MLIP with first-principles results (see Methods), as shown in Fig.~\ref{fig1}c.
We observed excellent agreement, with the largest deviation at the M-point. 

In summary, we demonstrate that a satisfactory Allegro MLIP for monolayer NbSe$_2$ in the high temperature phase, i.e., without CDWs, can be developed with relatively little training data, and we move on to investigating if Allegro can also capture the CDW phases.

\subsection{Monolayer - CDW phase}

\begin{figure*}
    \centering
    \includegraphics[width=\linewidth]{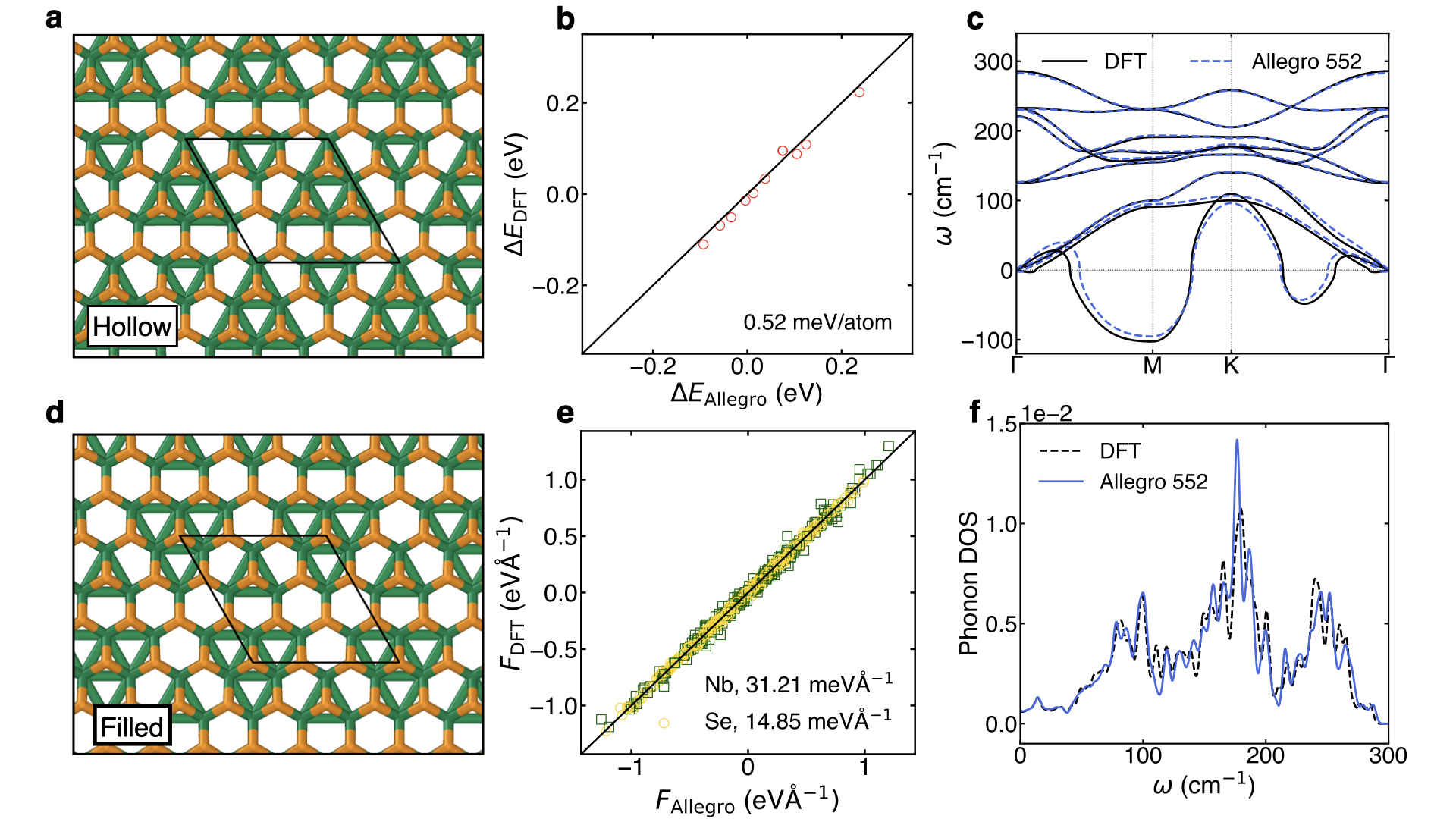}
    \caption{\textbf{Development of a MLIP for NbSe$_{2}$ in the normal and CDW phases.} 
    \textbf{a} - Structure of the hollow CDW phase of $\mathrm{NbSe_2}$, with the $3\times3$ supercell shown. CDW distortions in the Nb atom positions are highlighted with additional sticks between Nb atoms, using a cutoff of 3.45~\AA. 
    \textbf{b} - Energy MAE per simulation frame from MD simulations at 200~K in the NVT ensemble, comparing Allegro MLIP predictions with DFT. 
    \textbf{c} - Phonon dispersion curves for the normal state, computed using DFT (solid lines) and Allegro MLIP (dashed lines), demonstrating close agreement. 
    \textbf{d} - Structure of the filled CDW phase, shown for a $3\times3$ supercell with distortions highlighted as in panel a. 
    \textbf{e} - Force parity plot for Nb and Se atoms, comparing MLIP and DFT forces. 
    \textbf{f} - Phonon density of state, computed at $\Gamma$, for the hollow CDW phase, comparing DFT and MLIP. 
    Further details are provided in the SI.}
    \label{fig2}
\end{figure*}

Building on these results, we develop a model to study CDW phases under small-smearing conditions (low electronic temperatures). 
In this regime, the symmetry-breaking transition creates a partial electronic gap, driven by the competition between the energy cost associated with periodic lattice distortion and the gain in electronic energy, stabilizing the CDWs relative to the normal state~\cite{lian2018unveiling}. 
Fig.~\ref{fig2}a, d show the $3\times3$ CDW structures for the commonly observed `hollow' and `filled' phases, characterized by in-plane Nb displacements.
In the hollow phase, Nb atoms move toward the center of a hexagon, while in the filled phase they shift toward a Se atom.
Our DFT calculations indicate that these phases are, respectively, 40~meV and 35~meV lower in energy than the normal state, consistent with earlier findings \cite{lian2018unveiling}.
With 27 atoms per unit cell, the energy difference between the normal and each CDW phase corresponds to $\sim$1~meV/atom, while the energy difference between the CDW phases being $\sim$1/6~meV/atom, setting the energy accuracy required to distinguish these competing configurations.
\begin{figure*}
    \centering
    \includegraphics[width=0.99\linewidth]{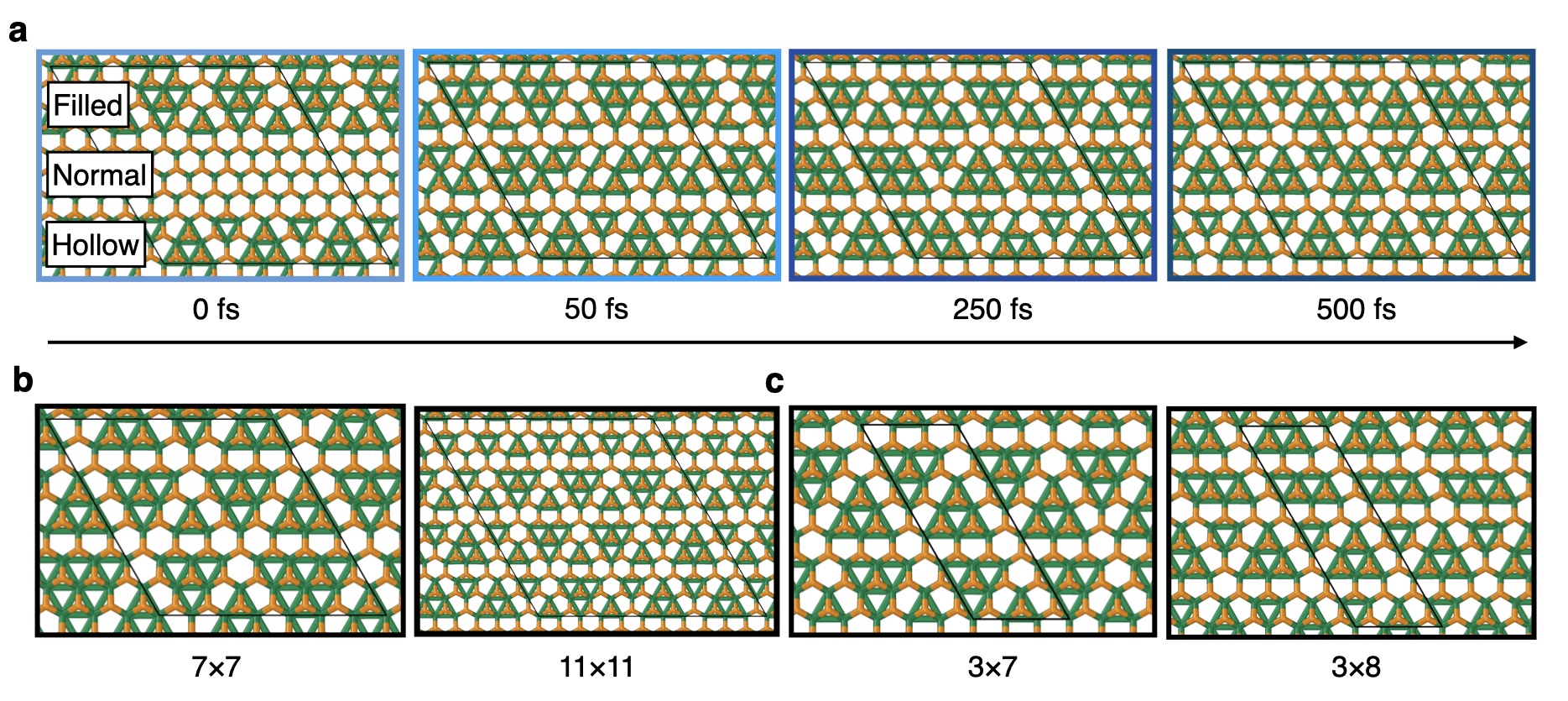}
    \caption{\textbf{Relaxations of commensurate and incommensurate supercells of NbSe$_2$ monolayers.} 
    \textbf{a} – MD NVT simulation at 10~K, starting from a $9\times9$ supercell initialized with a mixture of hollow, normal, and filled motifs. The system evolves toward the hollow CDW structure. 
    \textbf{b} – Relaxations of incommensurate $7\times7$ and $11\times11$ supercells. The $7\times7$ case shows mainly filled-type distortions with some hollow features, while the $11\times11$ case displays alternating hollow and filled motifs. 
    \textbf{c} – Incommensurate $3\times7$ and $3\times8$ supercells also show coexistence of hollow and filled patterns, with the $3\times8$ case exhibiting less well-developed distortions.}
    \label{fig3}
\end{figure*}

We initially trained MLIPs on 889 frames for the $3\times$3 monolayer with a single lattice parameter. Approximately half of these were generated via FLARE and half from relaxation after random displacements (details in the Methods).
The hyperparameter scan (see SI) identified $r_{\mathrm{max}} =5$~\AA, $l_{\mathrm{max}} = 5$, and $n_{\mathrm{layers}} = 2$ as reasonable values. 
Given that the CDW length scale is approximately 10.5 \AA, an $r_{\mathrm{max}}$ of 5~\AA ~captures practically all of the environments.
An $r_{\mathrm{max}}$ of 10 \AA~ also performs well, as shown in the SI, but intermediate values showed poorer results.

Testing this preliminary model on larger commensurate and incommensurate supercells revealed deviations in the lowest-energy structures and large errors in forces, showing a clear breakdown in extensibility.
This challenges the common assumption that MLIPs trained on smaller cells can be directly applied to larger supercells, underscoring the subtlety of CDWs compared to most applications, and the need for more extensive training in such systems. 
To address this, we expanded the dataset with 200 strained $3\times3$ frames ($\pm1$\%, $\pm2$\%) generated via FLARE, and 335 incommensurate supercells obtained from iterative training of previous models, bringing the total to 1424 structures for the monolayer dataset.
The incommensurate structures improved the MLIP's extensibility, but introduced rippling effects for larger supercells (e.g., $10\times10$), which were mitigated through the inclusion of the strained frames. Together, these additions enabled robust modeling of CDWs in NbSe$_2$ (see SI).

\begin{figure*}
\centering\includegraphics[width=0.99\linewidth]{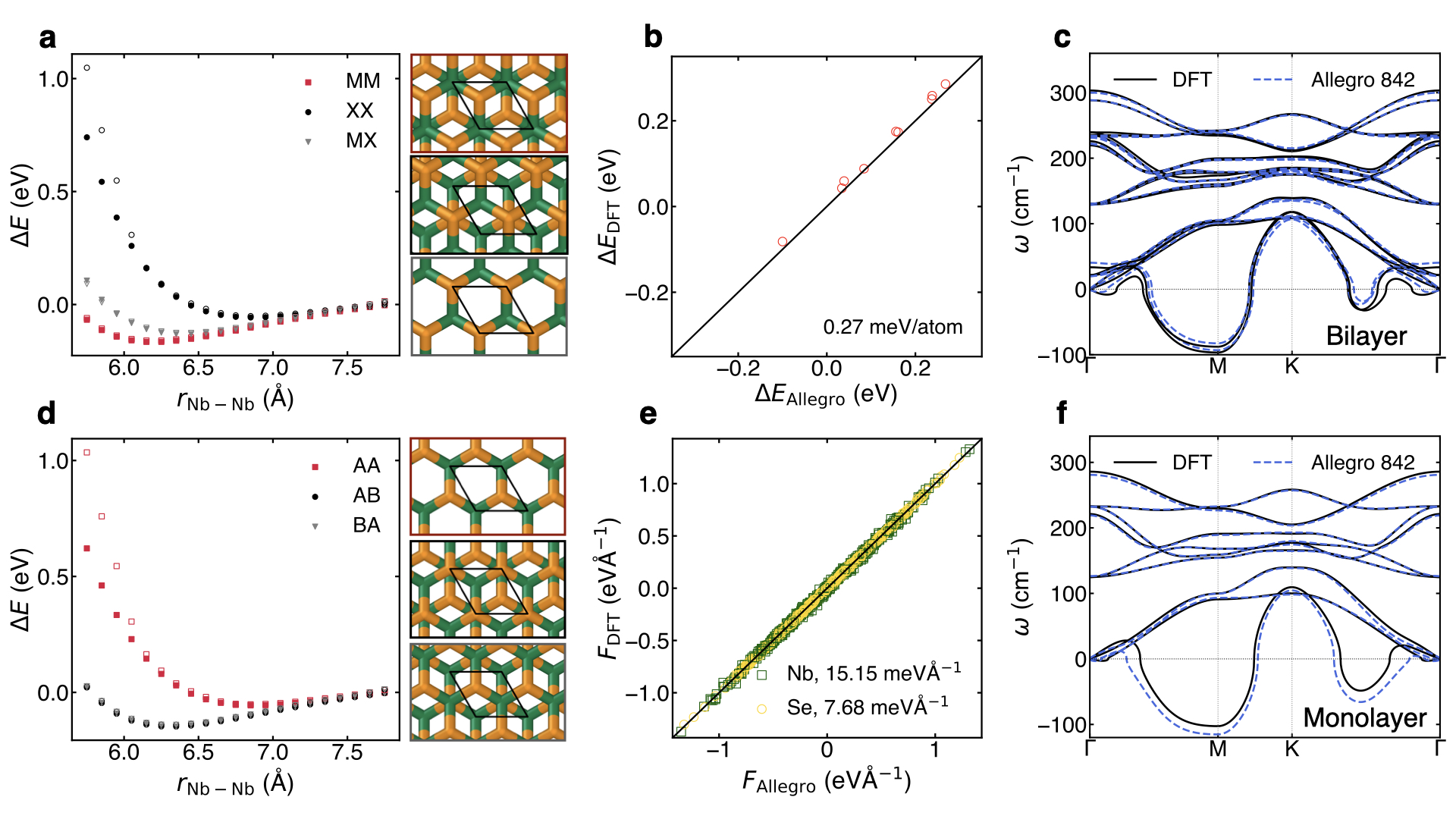}\caption{\textbf{Development of a MLIP for bilayer NbSe$_{2}$.} \textbf{a} - Binding energy curves for high-symmetry stackings of the 180\degree ~$\mathrm{NbSe_2}$ bilayer. Filled symbols represent Allegro MLIP results, and empty symbols represent DFT results.
\textbf{b} Energy MAE per simulation frame from MD simulations at 200~K in the NVT ensemble, comparing Allegro MLIP predictions with DFT. 
\textbf{c} - Phonon dispersion curves for the normal phase of the bilayer in the MM stacking. A comparison is shown between DFT (solid lines) and Allegro MLIP (dashed lines). 
\textbf{d} - Binding energy curves for high-symmetry stackings of the 0\degree ~$\mathrm{NbSe_2}$ bilayer. Filled symbols represent Allegro MLIP results, and empty symbols represent DFT results. 
\textbf{e} - Force parity plot for Nb and Se atoms, comparing MLIP and DFT forces. 
\textbf{f} -Phonon dispersions of the monolayer evaluated with the bilayer MLIP, compared with DFT. 
Further details are provided in the SI. }
    \label{fig4}
\end{figure*}

We validated the MLIP via MD simulations at 200~K in the NVT ensemble for a $3\times3$ structure, comparing energies and forces against DFT (Fig.~\ref{fig2}b, e). The model achieved a force MAE of 14.85~meV/\AA~for Se and 31.21~meV/\AA~for Nb (i.e., around $\sim$5\%), and an energy MAE of 0.52~meV/atom, which is below the energy threshold for distinguishing the normal and CDW phases, but above that required to differentiate the CDW phases.
To probe this further, we relaxed the normal, filled and hollow structures and found that, relative to the normal state, the hollow configuration is 32~meV lower in energy and the filled one 28~meV. This demonstrates that our model captures the relative energy differences between phases reasonably well: the energy MAE at 200~K reflects overall performance but is not sufficient to assess its ability to capture subtle CDW energy differences near local-minima and the ground-state.
Additional parity plots for incommensurate structures are provided in the SI, further illustrating the transferability of the model. 

Beyond energy and force validation, we also assessed dynamical stability by computing phonon dispersions for both the normal (1$\times$1) and hollow CDW (3$\times$3) phases, and comparing with DFT (Fig.~\ref{fig2}c, f). For the normal phase, phonons were evaluated in a 3$\times$3 supercell. 
While this supercell does not yield fully converged phonons—unlike in the large-smearing case—it enables a direct comparison under the same approximations as DFT. As expected, the normal phase exhibits an unstable acoustic phonon mode, indicative of a CDW instability, with negative frequencies around $\Gamma$-M and M-K \cite{bianco2020weak, zheng2019electron, calandra2009effect}; the latter is a spurious effect due to the supercell size, discussed in more detail later. 
Overall, the phonon dispersions agree well across the Brillouin zone, including the unstable mode. For the hollow phase (Fig.~\ref{fig2}f), due to the high density of modes, we report only the phonon density of states at $\Gamma$. In this case, the MLIP results are in close agreement with the DFT ones, consistent with the absence of unstable modes.

Since several phases exist in monolayer NbSe$_2$, we performed low-temperature MD at 10~K (NVT) to demonstrate how they coexist and interconvert.
We started from a configuration with coexisting hollow, normal, and filled regions: a $3\times3$ supercell of the CDW phases with the bottom $1\times3$ being hollow, middle $1\times3$ being normal and top $1\times3$ being filled. 
The time evolution is shown in Fig.~\ref{fig3}a.
After $\sim$50~fs, the initially normal regions develop CDW-like distortions resembling both hollow and filled phases, while the original hollow/filled domains remain stable.
By $\sim$250~fs, the former normal region predominantly exhibits hollow-like CDW distortions. 
Finally, after $\sim$450~fs the entire supercell is dominated by hollow-like CDW features, although the region that was initially filled retains some filled-like character.
At longer times, the system relaxes fully to the hollow phase, consistent with literature.
This confirms that, despite high-temperature validation errors ($\sim$0.5 meV/atom) exceeding the energy difference between CDWs ($\sim$0.2 meV/atom), the MLIP is sufficiently reliable at low temperature to reproduce their relative stability.

\begin{figure*}
     \centering\includegraphics[width=0.99\linewidth]{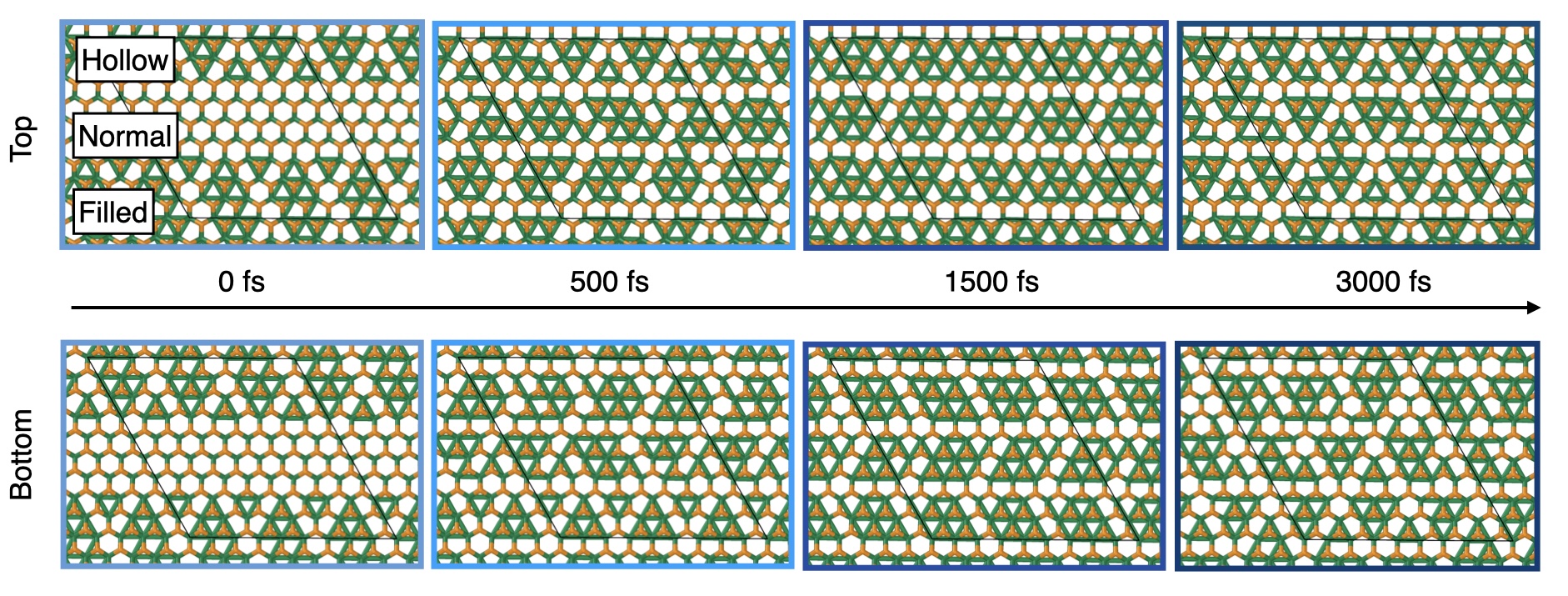}\caption{\textbf{Results for the MLIP applied to NbSe$_{2}$ bilayers.} MD NVT simulation at 10 K, starting from a 9$\times$9 supercell prepared with a mixture of hollow, normal, and filled phases in both layers. Intermediate frames illustrate the evolution of the system during the relaxation process, with the final structure corresponding to the stable configuration.}
     \label{fig5}
 \end{figure*}

So far, we have tested our model on commensurate supercells (e.g., $3\times$3 or  multiples thereof), which capture the CDW periodic lattice distortions.
However, in contexts such as moiré patterns or substrate-induced effects, incommensurate supercells are often used to reduce computational cost, but their boundary conditions complicate the interpretation of CDW behavior.
To illustrate this, in Fig.~\ref{fig3}b, we show relaxations for $7\times$7 and $11\times$11 supercells.
The $7\times$7 case exhibits predominantly filled-type distortions, with some hollow features also present, corresponding to the larger $2\times$2 triangular distortions seen in the pristine phases.
The $11\times$11 case displays alternating filled and hollow triangular distortions along certain directions, with mixed tessellation elsewhere.
Similar behavior is observed (Fig.~\ref{fig3}c) for incommensurate $3\times7$ and $3\times8$ supercells: filled and hollow patterns coexist, but in the $3\times8$ case the CDW is less well developed, with predominantly hollow motifs alternating with poorly formed distortions.
These results suggest that incommensurability may contribute to coexisting hollow and filled distortions, consistent with previous observations in twisted bilayers~\cite{Cheung2024}.

\subsection{Bilayer}

As monolayer NbSe$_2$ is rarely isolated experimentally, bilayers provide a more realistic platform for comparison with experimental measurements. ~\cite{holler2019air, cao2015quality, wang2017high, fu2024van}
To this end, we trained a model for bilayer NbSe$_2$ starting from a subset of the monolayer dataset consisting of 471 $3\times3$ frames generated by BAL, and subsequently added 1013 new bilayer frames spanning five stacking configurations with varying interlayer separations, each sampled at parallel ($0^{\circ}$) and antiparallel ($180^{\circ}$) orientations (BAL at 200~K). The final dataset therefore consists of 1484 frames in total.
In this case, hyperparameters scans identified $r_\mathrm{max}=8$~\AA, $l_\mathrm{max}=4$ and $n_\mathrm{layers}=2$ as reasonable values. 
More details about the dataset and hyperparameters are provided in the SI.
The primary adjustment of a larger $r_\mathrm{max}$ arises from the equilibrium interlayer distance between Nb atoms ($\approx$ 6.25~\AA), requiring $r_\mathrm{max}=8$~\AA~to accurately capture interlayer coupling.

To assess the bilayer MLIP, we followed the same validation strategy as for the monolayer.
The binding energy curves for our model and DFT (Fig.~\ref{fig4}a, d), for the indicated stackings, show good agreement, 
with minor discrepancies at extreme interlayer distances that were not thoroughly sampled by the BAL.
For the most stable (natural) MM stacking, energy and force comparisons from MD simulations at 200~K (NVT) confirm the high accuracy of the MLIP (Fig.~\ref{fig4}b, e). The bilayer model yields an energy MAE of 0.27~meV/atom and force MAEs of 15.15~meV\AA$^{-1}$ for Nb and 7.68~meV\AA$^{-1}$ for Se, values that are even smaller than in the monolayer.
We further investigated CDW stacking in bilayers. Among the nine possible filled–filled, hollow–hollow, and hollow–filled arrangements (not all unique, see SI), MLIP and DFT both identify the hollow–filled configuration as the lowest-energy state for anti-parallel MM stacking.

The phonon spectra (Fig.~\ref{fig4}c) for the natural stacking agree well with DFT, with only minor deviations in the unstable and high-energy optical modes.
The bilayer spectrum essentially duplicates that of the monolayer, with each mode is nearly doubly degenerate due to weak interlayer coupling.
As in the monolayer case, these spectra were computed using a $3\times3$ supercell as a proof of concept; full convergence would require larger cells and additional training refinement, which we discuss in a later section.
Importantly, the developed MLIP captures both monolayer (Fig.~\ref{fig4}f) and bilayer phonons consistently using the same dataset and hyperparameters.

We then explored the coexistence of CDWs in bilayers by initializing a $9\times9$ supercell with normal, hollow, and filled distortions in each layer (Fig.~\ref{fig5}). At 10~K, the system consistently evolved into a mixed configuration, with one layer adopting a filled CDW and the other a hollow CDW. This arrangement is the low-energy stacking identified by our DFT calculations and others \cite{Cheung2024}. While the hollow layer remains nearly pristine, the filled layer shows deviations from the ideal distortion, suggesting that the hollow motif is more stable and that at 10~K the filled phase is near its transition temperature. Importantly, direct relaxation of these mixed states without annealing often yields local minima rather than the true ground state, underscoring the critical role of annealing in CDW reconstruction for bilayers and, more broadly, for complex layered systems with CDWs.

\section{Phonons and transition temperature}

\begin{figure*}
    \centering
    \includegraphics[width=0.99\linewidth]{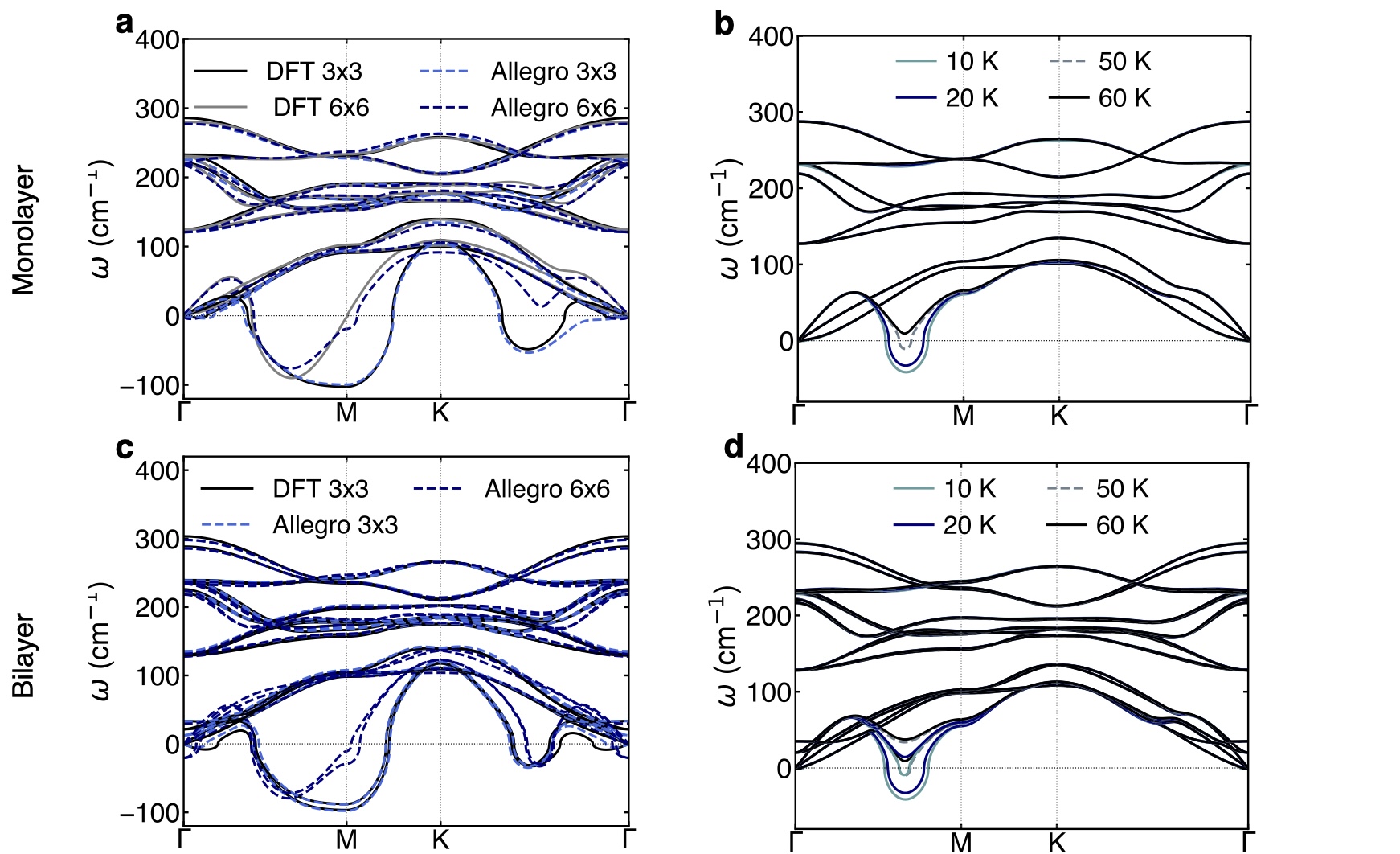}
    \caption{\textbf{Phonon dispersions and SSCHA results from the refined datasets.} 
    \textbf{a} –  Monolayer phonon dispersions from DFT (solid lines) and MLIP (dashed lines) using $3\times3$ and $6\times6$ supercells with Phonopy. Larger cells suppress the spurious M–K instability and shift the soft-mode minimum to $q_\mathrm{CDW}=2/3\,\Gamma$–M. 
    \textbf{b} – Temperature-dependent phonons from SSCHA for the monolayer ($10$–$60$~K), showing progressive softening and eventual instability of the CDW mode. 
    \textbf{c} – Bilayer (MM stacking) phonon dispersions from DFT ($3\times3$ supercell, solid lines) and MLIP (dashed lines).
    \textbf{d} – Bilayer SSCHA results ($10$–$60$~K), indicating a $T_\mathrm{CDW}$ close to the monolayer, with only a slight reduction.}
    \label{fig6}
\end{figure*}

Beyond structural reconstructions, accurately capturing phonons and related properties, such as transition temperatures ($T_{\mathrm{CDW}}$), remains challenging.
Unlike structures, which depend primarily on local bonding, phonons—especially the soft acoustic mode driving the CDW—require accurate long-wavelength information and well-converged interatomic force constants. Small supercells (e.g., $3\times3$) reproduce local energies and forces but fail to capture these extended correlations, leading to spurious instabilities and preventing full convergence. Thus, unlike many MLIP applications, simple extrapolation from short periodicity does not work for CDWs. 

This limitation is illustrated in Supplementary Fig. 22.
While phonon dispersions of the monolayer obtained from our MLIP appear largely size-independent, DFT results show pronounced dependence on the supercell size: instabilities disappear in larger supercells and M-point phonons become positive, consistent with previous calculations \cite{bianco2020weak}.
This reveals a key weakness of the developed MLIP. Although trained on a mix of supercells, the dataset was dominated by $3\times3$ structures and restricted by a short real-space cutoff ($r_\mathrm{max}$), roughly one CDW wavelength. As a result, the model fails to capture the long-range interatomic force constants that govern the soft acoustic mode, limiting phonon transferability across supercells in training and reducing extensibility to larger ones.

To overcome this and refine the monolayer dataset for vibrational properties, we adopted a three-step strategy.
First, we expanded the dataset with 246 MD snapshots at 200~K of $4\times4$–$7\times7$ and selected rectangular ($3\times n$) supercells, which yielded only marginal improvements (see SI). Second, to reduce the bias toward small cells, we pruned the overrepresented $3\times3$ structures while retaining the larger ones, producing a more balanced dataset. 
Third, we increased the cutoff to $r_\mathrm{max}=10$~\AA, enabling the model to capture longer-range force constants. The final dataset comprised 462 structures: 354 square cells (including $3\times3$–$8\times8$) and 108 rectangular cases ($3\times2$–$3\times8$). A complete breakdown and hyperparameter scans are provided in the SI.

These refinements markedly improved the phonon spectra. As shown in Fig.~\ref{fig6}a, larger cells shift the instability minimum to $q_\mathrm{CDW}=2/3,\Gamma$–M (corresponding to the $3\times3$ distortion) and suppress the spurious M–K instability, consistent with DFT. Increasing cell size beyond $6\times6$ slightly worsens quantitative agreement, indicating that full convergence (e.g., $12\times12$) would require significantly more large-cell training data \cite{bianco2020weak, benić2025machine}, and thus substantially higher computational cost. However, rather than pursuing exhaustive convergence, our focus is on training protocols that enable reliable structural reconstruction and provide general guidelines for dataset design in CDW systems, while still capturing the essential phonon physics.

Building on these monolayer refinements, we next turned to bilayers, aiming to apply the same strategy. Starting from the pruned monolayer dataset, we tested variants that incorporated either the full bilayer set previously used (1013 frames) or reduced subsets of $3\times3$ bilayer structures (150–250 frames). 
The reduced subsets proved insufficient: they failed to correct the spurious instability introduced by the bias toward smaller structures and gave a worse overall description of the spectra, as well as discrepancies in binding energy curves (see SI).
By contrast, combining the pruned monolayer dataset with the full bilayer set and extending the cutoff  did not completely eliminate this bias, but did improve the description of the main phonon instability, as shown in Fig.~\ref{fig6}c. 
These findings highlight a key bottleneck: accurate bilayer phonons require training on larger bilayer supercells, which are prohibitively expensive at the DFT level. Nevertheless, the trends confirm that the same guidelines established for monolayers—exposure to larger supercells and longer cutoffs—must also be followed for bilayers and more complex systems.

Beyond phonon spectra, MLIPs can also be used to estimate CDW transition temperatures, which requires accurately capturing anharmonic effects, subtle energy shifts due to structural distortions, and vibrational excitations.
Classical MD alone cannot fully describe vibrational entropy contributions arising from quantized fluctuations \cite{bianco2019quantum, bianco2020weak}.
To overcome this, we combined our Allegro-MLIP with the stochastic self-consistent harmonic approximation (SSCHA) \cite{monacelli2021stochastic, bianco2017second} to compute temperature-dependent phonon dispersions.

For the monolayer, SSCHA shows that the soft mode at $q_\mathrm{CDW}=2/3,\Gamma$–M softens near 60 K and becomes unstable at 50 K (Fig.~\ref{fig6}b), placing $T_\mathrm{CDW}$ in the 50–60 K range. This estimate is slightly below earlier SSCHA work (73 K \cite{bianco2020weak}) but in close agreement with STM \cite{arguello2014visualizing}. Classical MD, in contrast, predicts lower values (see SI), reflecting the role of quantum fluctuations.

For the MM bilayer stacking, SSCHA places $T_\mathrm{CDW}$ in the 50–60 K range (Fig.~\ref{fig6}d), very close to the monolayer estimate and consistent with the weak dimensionality dependence reported previously \cite{bianco2020weak}. Notably, the mixed hollow/filled bilayer configuration is stable only up to $\sim$5 K, beyond which the filled layer undergoes dynamic fluctuations (Fig.~\ref{fig5}b and SI, where a classical estimate for the $T_\mathrm{CDW}$ is shown). Additional AB stacking results (see SI) show CDWs suppressed at lower temperatures than MM, underscoring stacking sensitivity.

Because SSCHA depends  on the  forces driving its temperature evolution, predictions of $T_\mathrm{CDW}$ are  sensitive to the underlying exchange–correlation functional, computational setup, and structural details \cite{fumega2023anharmonicity, libbi2024ultrafast}, and—in the MLIP case—to the fidelity of the potential itself. In NbSe$2$, where CDW and normal states differ by only meV/atom, this sensitivity is amplified. It helps explain small discrepancies between our SSCHA results and prior literature, and cautions that even the close agreement between our monolayer and bilayer $T_\mathrm{CDW}$ values must be interpreted carefully. In practice, the monolayer MLIP provides the most reliable absolute estimates, while the bilayer MLIP—more limited by available data—gives nearly identical values with a slight reduction. Tests with a unified MLIP (see SI) consistently place the T$_\mathrm{CDW}$ of the bilayer slightly below that of the monolayer, providing further support for the weak dimensionality dependence reported in the literature.

Dataset refinements proved essential for improving phonon spectra and $T_\mathrm{CDW}$ estimates but came at the cost of reduced structural accuracy—most notably in bilayers, where the limited number of large-cell training structures made it impossible to simultaneously mitigate the small-cell bias and fully capture interlayer coupling. For this reason, we retain two complementary MLIPs: one optimized for structural reconstruction and one refined for vibrational properties. This dual approach balances accuracy across properties while providing guidelines for future training. Based on our analysis, a unified monolayer–bilayer potential capable of simultaneously capturing both structural and vibrational CDW physics will likely require substantially larger, carefully balanced datasets on the order of a few thousand frames.

\section{Conclusions}

Here we investigated protocols for training robust and transferable MLIPs for CDWs in NbSe$_2$, with applicability to other layered quantum materials. We found CDW periodic lattice distortions are straightforward to learn, vibrational properties are more demanding, as we found limited transferability of our models.

For monolayers, accurate MLIPs for structural reconstruction can be obtained from modest datasets ($\sim$ 600–1000 frames) combining BAL with relaxations, with the desired transferability is achieved through including strained and incommensurate structures. At low temperature, we consistently predict the hollow CDW as the most stable phase, in agreement with STM and DFT studies. For bilayers, dataset balancing and larger supercells remain essential, but the associated DFT cost represents a key bottleneck. In the natural stacking, the hollow–filled arrangement emerges as the lowest-energy configuration, though the details of CDW reconstruction and relative energetics remain sensitive to stacking.

Accurate vibrational properties require larger supercells, extended cutoffs, and carefully curated datasets. To address this, we developed a refined MLIP tailored to vibrational properties, complementing the structural model and providing a framework for systematic improvement. Using the refined vibrational model, we combined Allegro with SSCHA to examine phonons and transition temperatures. For monolayers, $T_\mathrm{CDW}$ is predicted to be $\sim$60 K, close to STM measurements and DFT studies. Bilayers yield very similar estimates, consistent with the weak dimensionality dependence reported previously. 

Overall, this work provides a clear workflow for training MLIPs tailored to different objectives—structural reconstruction versus vibrational properties—across monolayers and bilayers. This framework enables the systematic study of CDW evolution with layer number and stacking, with direct relevance to multilayer and twisted systems. Future extensions, including three-layer systems, doping, and substrate interactions, will broaden applicability to increasingly complex experimental settings.

\section{Methods}\label{methods}

We combined first-principles density functional theory (DFT) with machine-learned interatomic potentials (MLIPs), in the Allegro architecture~\cite{musaelian2023learning},  to model structural and vibrational properties of NbSe$_2$. Training datasets were generated through Bayesian active learning (BAL) and relaxations from random displacements. The trained Allegro models were integrated into LAMMPS~\cite{thompson2022lammps} for molecular dynamics (MD), structural relaxations, binding energy calculations, among other calculations. Phonons were computed using Phonopy~\cite{togo2023first} in combination with the MLIPs, and finite-temperature phonon properties were obtained using the stochastic self-consistent harmonic approximation (SSCHA)~\cite{monacelli2021stochastic, bianco2017second}. Dataset composition, hyperparameters, and additional convergence tests are reported in the Supplementary Information (SI).

\subsection{DFT calculations}

All DFT calculations were performed with Quantum ESPRESSO~\cite{giannozzi2009quantum, giannozzi2017advanced} using ultrasoft pseudopotentials within the vdW-DF2-c09 exchange--correlation functional~\cite{cooper2010van}. The plane-wave cutoff energies were 70~Ry for wavefunctions and 560~Ry for charge density. A $27\times27\times1$ Monkhorst--Pack $k$-point grid was used for the primitive three-atom cell and rescaled accordingly for larger supercells to keep the same $k$-point-spacing. Smearing was treated with the Methfessel--Paxton scheme, with a Gaussian broadening of 0.005~Ry (small smearing, CDW-stabilized case) or 0.05~Ry (large smearing, normal state). All systems were treated as non-magnetic. The Coulomb cutoff~\cite{sohier2017density} was used with 40~\AA{} vacuum spacing.

\subsection{Dataset generation}\label{dataset}

Training data was generated using two complementary approaches. BAL was carried out with the FLARE framework~\cite{vandermause2020fly}, which employs Gaussian process regression on atomic cluster expansion descriptors to adaptively select configurations for DFT evaluation during MD simulations. In parallel, we generated additional structures by applying random Gaussian displacements of width $\sigma=0.1$~\AA{} to equilibrium configurations, followed by DFT relaxations. To reduce how correlated the frames were, only a subset of frames was retained (every third frame, after the third). Together, these methods provided diverse configurations spanning both equilibrium and non-equilibrium structures. Details of frame selection, DFT settings during BAL, and convergence checks are reported in the SI.

\subsection{Machine-learning potentials}\label{ML}

MLIPs were trained using the Allegro equivariant neural network architecture~\cite{musaelian2023learning}, which combines strict locality with tensor equivariance for data efficiency and scalability. We used an 80/20 split between training and validation data.
Angular order, cutoff radius, and number of tensor layers were tuned in a hyperparameter search, details of which can be found in the main text and SI. 
Complete training parameters and validation metrics are provided in the SI.

The trained MLIPs were interfaced with LAMMPS for MD (NVT ensemble, Nosé--Hoover thermostat, 0.5~fs timestep), structural relaxations (conjugate gradient minimization), and binding energy curve calculations. Validation was performed by sampling MLIP-generated structures and recomputing their energies and forces with DFT.

\subsection{Phonon and SSCHA calculations}

Phonon dispersions were computed with finite-differences using Phonopy \cite{togo2023first}, coupled either to DFT or the trained MLIPs. Supercells used for phonons were matched across DFT and MLIP calculations for consistency. Acoustic sum-rule corrections were applied as implemented in Phonopy.

Finite-temperature phonon properties were obtained using the SSCHA formalism~\cite{monacelli2021stochastic, bianco2017second}, with MLIP-derived dynamical matrices as input and MLIP forces used during variational sampling. Accurate description of the CDW soft mode required larger supercells of $6\times6$ for both monolayers and bilayers. SSCHA simulations were carried out in the NVT ensemble over 0--70~K ( for both mono- and bilayer), with $\sim$2000 stochastic configurations per temperature to ensure convergence. Further numerical details and convergence tests are reported in the SI.

\section*{Acknowledgments}

We thank Cameron Owen and Stefano Falletta for stimulating discussions. N.R acknowledges support through the the Enterprise Science Fund (Project 218056—  Twisted NbSe$_2$). Computational resources are provided for by Harvard FAS Research Computing. Z.A.H.G acknowledges support through the Glasstone Research Fellowship in Materials, University of Oxford. F.L. acknowledges support through the Swiss National Science Foundation (SNSF) mobility fellowship P500PT\_217861. A.O.F and J.L.L. acknowledge the computational resources provided by the Aalto Science-IT project and the financial support from Finnish Quantum Flagship, InstituteQ, the Jane and Aatos Erkko Foundation and the Academy of Finland Projects Nos. 331342, 358088, 349696 and 369367.  C. T. S. C. acknowledges funding from the Croucher Foundation and an Imperial College President's Scholarship. This work used the ARCHER2 UK National Supercomputing Service (https://www.archer2.ac.uk) and the Imperial College Research Computing Service facility CX1 (DOI: 10.14469/hpc/2232). We acknowledge the Thomas Young Centre under Grant No. TYC-101.

\bibliography{main}
\end{document}